\documentclass[conference]{IEEEtran}
\usepackage{blindtext, graphicx}
\usepackage{lipsum}
\usepackage{amsmath}
\IEEEoverridecommandlockouts

\ifCLASSINFOpdf
\else
\fi
\begin{document}
%
\title{A Minimum-Cost Flow Model for Workload Optimization on Cloud Infrastructure}

\author{\IEEEauthorblockN{Frederick Nwanganga, Mandana Saebi, Gregory Madey and Nitesh Chawla \thanks{This research was supported in part by the National Science Foundation (NSF) Grant IIS-1447795.}}

\IEEEauthorblockA{Department of Computer Science and Engineering\\University of Notre Dame \\Notre Dame, IN 46556\\\{fnwangan, msaebi, gmadey, nchawla\}@nd.edu}}


%


\maketitle
\thispagestyle{plain}
\pagestyle{plain}

\setcounter{page}{1}
\pagenumbering{arabic}

\begin{abstract}
Recent technology advancements in the areas of compute, storage and networking, along with the increased demand for organizations to cut costs while remaining responsive to increasing service demands have led to the growth in the adoption of cloud computing services. Cloud services provide the promise of improved agility, resiliency, scalability and a lowered Total Cost of Ownership (TCO). This research introduces a framework for minimizing cost and maximizing resource utilization by using an Integer Linear Programming (ILP) approach to optimize the assignment of workloads to servers on Amazon Web Services (AWS) cloud infrastructure. The model is based on the classical minimum-cost flow model, known as the assignment model.
\end{abstract}



%
\IEEEpeerreviewmaketitle

\section{Introduction}

With the increase in globally accessible, high speed connectivity; the commoditization of processing and storage; and the increased ubiquity of server virtualization, enterprise technology trends are moving from monolithic industrial automation to ``globally accessible infrastructure that incorporates highly scalable, internet-centric cloud services delivered by external suppliers'' \cite{giga-cloud-arch}. As a result of this, organizations are increasingly starting to look at ways to leverage cloud services not only to meet the demands of their industry but also as a path towards relevance and survival. A survey of 502 IT decision makers at companies with 500 or more employees from different industries, showed that cost, scalability and business agility/responsiveness to change were the top three business drivers for the adoption of cloud service at these organizations \cite{giga-cloud-bench}. Cloud services, in general offer organizations a path to cost reduction and the avoidance of the backward inertia that comes with maintaining existing infrastructure. 

It is estimated that the public IaaS (Infrastructure-as-a-Service) market will grow from about \$23 billion to over \$34 billion by 2016 \cite{giga-cloud-outlook}. Given this projection, it is not surprising that many educational institutions have also started to adopt cloud computing to fulfill their various needs - from cloud based teaching and learning environments \cite{nku-cloud-solution, aws-classroom} to public or hybrid cloud university data centers. 

\section{Problem Definition \& Importance}

The flexibility and ease with which secure and scalable compute and storage resources can be provisioned and de-provisioned on-demand within the IaaS paradigm makes it an ideal, strategic and cost-effective option for most organizations to adopt. Due to its consistently high ratings in the areas of completeness of vision and ability to execute \cite{gartner-iaas-quadrant}, Amazon Web Services (AWS) is a leader in the IaaS space and the platform on which our work is based.

AWS provides a suite of Elastic Compute Cloud (EC2) instance types for different use cases. These instance types provide varying combinations of CPU, memory, storage and network capacity \cite{aws-ec2-instance-types}. In the course of a cloud implementation or migration effort, as services are identified for migration, users have the flexibility to choose the EC2 instance type that provides the appropriate mix of resources for the target application and workload. Currently, the choice of an instance type is usually based on a heuristic approach and does not guarantee that an optimal solution is selected with regards to performance and cost. In this paper, we present the use of an Integer Linear Programming (ILP) model for an efficient assignment of workloads to servers in order to reduce cost as well as to maximize resource utilization.

The rest of the paper is organized as follows. Section III presents some of the related work. Section IV introduces two AWS services: (1) Elastic Compute Cloud (EC2) and (2) Cloud Watch. In section V, we present our optimization methodology, the algebraic formulation for the model as well as the implementation of the model using AMPL \cite{ampl}. In sections VI and VII, we discuss our results and conclusions, respectively.

\section{Related Work}
Along with all of its advantages, cloud hosted infrastructure introduces a new set of challenges. With the increased flexibility and the large amount of metrics available, it becomes increasingly difficult and yet important to optimize the distribution of workloads across available resources. For organizations trying to determine the best path towards migrating their physical on-premise infrastructure to virtual cloud infrastructure, determining the optimal mapping of virtual machines to physical machines in a cloud data center is a very important optimization problem. It's a problem with a huge impact on cost, application performance, and energy consumption \cite{kokkinos2015sumo}, \cite{bartok2015branch}. 


The most similar work to ours is \cite{kokkinos2015sumo}, in which the authors developed a toolkit for analyzing and providing resource optimization suggestions for users of Amazon EC2 instances. Their open-source toolkit, SuMo, implements important functionality for collecting monitoring data from Amazon Web Services (AWS), analyzing them and suggesting changes that optimize the use of resources and the associated costs. SuMo consists of three main components/modules: \textit{cloudData}, which is responsible for collecting monitoring data, \textit{cloudKeeping}, which contains a set of Key Performance Indicators (KPI), and \textit{cloudForce}, which incorporates a set of analytic and optimization algorithms. Optimization in SuMo is performed using an ILP-based Cost and Utilization Optimization (CUO) mechanism that maximizes the utilization of the resources/instances and minimizes the cost of their use. Unlike this research, our work focuses more on the optimization task and considers a broader set of parameters (memory, disk IO and network IO) and constraints. 
 
Another work \cite{zhang2014dynamic}, with a similar objective but different approach, studies resource allocation in a cloud market through the auction of Virtual Machine (VM) instances. It first proposes a cooperative primal dual approximation algorithm with approximation ratio close to 2.72. Employing the cooperative approximation algorithm as a building block, it then designs a novel randomized auction using a pair of tailed primal and dual LPs (Linear Programs) to decompose an optimal fractional solution into a summation of a series of weighted valid integer solutions. 

Bart\'{o}k and Mann propose a custom branch-and-bound algorithm that exploits problem-specific knowledge to improve effectiveness \cite{bartok2015branch}. Based on their empirical results, they argue that the new algorithm performs better than state-of-the-art general-purpose ILP solvers. 

In \cite{nasirianiusing} a more detailed work on the analysis the usage of burstable instance is presented.  Using measurements from both Amazon EC2 and Google Compute Engine (GCE), they identified key idiosyncrasies of resource capacity dynamism for burstable instances that set them apart from other instance types. The network bandwidth and CPU capacity for these instances were found to be regulated by deterministic, token bucket like mechanisms. The paper discusses two case studies of how certain memcached workloads might utilize their modeling for cost-efficacy on EC2 based on: (i) temporal multiplexing of multiple burstable instances to achieve the CPU or network bandwidth (and thereby throughput) equivalent of a more expensive regular EC2 instance, and (ii) augmenting cheap but low availability in-memory storage offered by spot instances with backup of popular content on burstable instances.

Another efficient toolkit for modeling and simulation of cloud computing environments and evaluation of resource provisioning algorithms is developed in \cite{calheiros2011cloudsim}. It is an extensible simulation toolkit that enables modeling and simulation of cloud computing systems and application provisioning environments. The toolkit supports both system and behavior modeling of Cloud system components such as data centers, virtual machines and resource provisioning policies. Moreover, it exposes custom interfaces for implementing policies and provisioning techniques for allocation of VMs under inter-networked Cloud computing scenarios. Several researchers from organizations, such as HP Labs in U.S.A., are using this toolkit currently.

\cite{hao2014online} proposes a generalized resource placement methodology for online resource placement in a cloud system. The methodology proposed can work across different cloud architectures and resource request constraints, where the arrivals and departures of the requests are in real-time. The proposed algorithms are online in the sense that allocations are made without any knowledge of future resource requests. 

A lot of work has also been done with regards to energy efficient cloud resource allocation in \cite{beloglazov2010energy}, \cite{beloglazov2012energy} and \cite{goudarzi2012sla}. In \cite{beloglazov2010energy} the authors present a decentralized architecture for energy aware resource management for Cloud data centers. They define the problem in terms of minimizing energy consumption while simultaneously meeting QoS (Quality of Service) requirements and stated policies for VM allocation. The paper proposes three stages for the continuous optimization of VM placement and presents heuristics for a simplified version of the first stage. An architectural framework and principles for energy-efficient cloud computing is presented in \cite{beloglazov2012energy}. The paper discusses open research challenges, and resource provisioning and allocation algorithms for energy-efficient management of cloud computing environments. A study in \cite{berral2010integer} exposes how to represent a grid data-center based scheduling problem, taking advantage of virtualization and consolidation techniques, and formulating it as a linear integer programming problem. \cite{bellur2010optimal} presents an optimal technique to map virtual machines to physical machines such that the number of required PMs is minimized; which leads to reduced power consumption. It provides two approaches based on linear programming and quadratic programming techniques that improve over the existing theoretical bounds and efficiently solves the problem of VM placement in data centers.

The trade-off between the total energy cost and client satisfaction in a system is explored in \cite{goudarzi2012sla}. In this paper, a resource allocation problem aims to minimize the total energy cost of a cloud computing system while meeting the specified client-level SLAs in a probabilistic sense. In the proposed scheme, the cloud computing system pays a penalty for the percentage of a client’s requests that do not meet a specified upper bound on their service time. An efficient heuristic algorithm based on convex optimization and dynamic programming is presented to solve the aforesaid resource allocation problem.

Another similar work that considers SLA as a constraint, is \cite{ishii2011elastic}. In this work, an elastic-stream system that dynamically allocates computational resources on the cloud is presented. To minimize the charges for using the cloud environment while satisfying the SLA, the paper formulates a linear programming problem to optimize cost as a trade-off between an application’s latency and cost. The paper claims that its proposed approach could save 80\% of the costs of using Amazon EC2 while maintaining an application’s latency in comparison to a naive approach.

There are numerous works in cloud data analysis such as \cite{abadi2009data}, \cite{ostermann2009performance}, \cite{wang2010impact} and \cite{ward2012semantic}. In \cite{ostermann2009performance} an evaluation of the usefulness of the current cloud computing services for scientific computing is presented. The authors analyze the performance of the Amazon EC2 platform using micro-benchmarks and kernels. Guohui and Eugene \cite{wang2010impact} present a measurement study to characterize the impact of virtualization on the networking performance of the Amazon EC2 data center. They measure processor sharing, packet delay, TCP/UDP throughput and packet loss among Amazon EC2 virtual machines.

\section{Amazon Web Services}
\subsection{Elastic Compute Cloud}
Elastic Compute Cloud (EC2) is Amazon's resizable compute service in the cloud. It makes it very easy to launch compute resources with a variety of operating systems configured to the needs of the user. An EC2 instance can be launched by simply selecting a pre-configured Amazon Machine Image (AMI), configuring security, network access and a host of other features using the AWS web services API or a variety of management tools provided. EC2 instances come in three different purchasing models \cite{aws-ec2-purchasing}:
\begin{itemize}
\item \textbf{On-Demand instances}\\
On-demand instances let users pay for compute capacity by the hour. They don't require upfront payments or come with long-term contracts. On-demand instances can be increased or decreased at any time and are only billed based on usage. However, during periods of high demand, it is possible that users may not be able to launch on-demand instances in certain availability zones.
\item \textbf{Reserved instances}\\
Reserved instances are useful for applications that have steady state needs and have predictable usage. They provide significant cost savings compared to on-demand instances but usually require upfront payments and come with long-term contracts.
\item \textbf{Spot instances}\\
Spot instances allow users to bid on a maximum amount they are willing to pay for a particular instance type. Spot prices fluctuate based on supply and demand. When the spot price of a particular instance rises above the maximum set by the user, the instance is terminated by Amazon EC2. With this in mind, spot instances are best for applications with flexible start and end times.
\item \textbf{Dedicated Instances}\\
Dedicated instances are Amazon EC2 instances that run in a virtual private cloud (VPC) on hardware that's dedicated to a single customer. Dedicated instances are physically isolated at the host hardware level from instances that belong to other AWS accounts. Dedicated instances may share hardware with other instances from the same AWS account that are not Dedicated instances.

\end{itemize}

For our research, we limit ourselves to evaluating on-demand EC2 instances in order to independently assess the immediate cost impact of our instance type selections over a defined period of time.

\subsection{Cloud Watch}
CloudWatch \cite{aws-cloudwatch} is a service for monitoring AWS resources such as Amazon EC2 instances, Amazon Elastic Block Store (EBS) volumes, Elastic Load Balancers, and Amazon Relational Database Service (RDS) instances. Metrics such as CPU utilization, latency, and request counts are provided automatically for these AWS resources. CloudWatch can also be used to set alarms, automatically react to changes as well as collect custom metrics and logs generated by applications and services in AWS. 

Amazon EC2 instances send monitoring data to CloudWatch every one or five minutes, depending on the configuration. The EC2 CloudWatch metrics include \textit{CPUCreditUsage}, \textit{CPUCreditBalance}, \textit{CPUUtilization}, \textit{DiskReadOps}, \textit{DiskWriteOps}, \textit{DiskReadBytes}, \textit{DiskWriteBytes}, \textit{NetworkIn}, \textit{NetworkOut}, \textit{NetworkPacketsIn}, \textit{NetworkPacketsOut}, \textit{StatusCheckFailed}, \textit{StatusCheckFailed\_Instance}, \textit{StatusCheckFailed\_System} \cite{aws-cloudwatch-ec2-metrics}. 

Amazon CloudWatch also allows for the production and consumption of custom metrics from EC2 instances. For Linux-based instances, a locally installed monitoring perl script can be used to collect memory, swap, and disk space utilization data, which is then remotely reported to CloudWatch as custom metrics \cite{aws-cloudwatch-linux-custom}. A similar process is available for Windows-based systems where EC2Config can be used to collect and send memory and disk metrics to CloudWatch Logs \cite{aws-cloudwatch-windows-custom}.

\subsection{Data}
For our project, we curated AWS CloudWatch metrics collected over 6 months from 108 Windows servers running individually unique workloads. The servers are running in the University of Notre Dame's Amazon Virtual Private Cloud (VPC) and have workload characteristics which are specific to the application or service running on that server.

Our work makes use of the \textit{CPUUtilization} standard metric, which is a measure of the percentage of allocated EC2 compute units that are currently in use on any particular instance. It is the basis for our instance CPU attribute which we will discuss later. We also make use of the \textit{MemoryUtilization} custom metric for our work. Like the CPU metric, this metric is the measure of the percentage of allocated memory utilized by an instance.

\section{Optimization}
Our research presents a model for cost and utilization optimization for servers running in AWS. By identifying underutilized or under-provisioned resource capacity, we can highlight cost reduction or performance improvement opportunities within our current set of running instances.

\subsection{Methodology}
Our optimization model is an Integer Linear Programming (ILP) model formulated as a modification of the classic assignment problem. In our model, the origins (or sources) are represented by a set of EC2 servers currently running in AWS, while the destinations are represented by the set of all publicly available EC2 instances. The optimization problem, therefore, is to assign all instances in the source to the instances in the destination, while minimizing cost and subject to the constraint that the host instances have enough resource capacity to handle the observed resource demand of the source instances.

Each instance in the model is defined by four attributes:
\begin{itemize}
\item \textbf{CPU}\\
The historical average CPU demand (plus two standard deviations) for the instance based on the metrics retrieved from Amazon CloudWatch. The CloudWatch CPU metrics are defined as a percent of the CPU capacity of the running instance. For our project, we re-factor this metric against the published Elastic Compute Unit (ECU) capacity for the instance type.
\item \textbf{Memory}\\
The historical average Memory demand (plus two standard deviations) for the instance based on the metrics retrieved from Amazon CloudWatch. The CloudWatch memory metrics are defined as a percent of the memory capacity of the running instance. For our project, we re-factor the metric against the published memory capacity (in GiB) for the instance type.
\item \textbf{Type}\\
The current EC2 instance type of the running server. This attribute includes the type of operating system, the instance model and the region in which the instance is running. For example a general purpose m4.large server running Red Hat Enterprise Linux (RHEL) in the US East Region (N. Virginia) will be represented as \texttt{rhel.m4.large.us-east}.
\item \textbf{Cost}\\
The On-demand Hourly Cost for each running EC2 instance (as listed in \cite{aws-ec2-purchasing}). On-demand costs are specific to the type of operating system, the instance model and the region in which an instance is running.
\end{itemize}

\subsection{Algebraic Formulation}
As described in the ``Optimization'' section, our model receives as input a set of running instances and outputs a recommendation for the allocation of the instances across all publicly available EC2 instance types. We denote the set of running instances as: 
\begin{align}
    S = \{I_1,I_2,...,I_M\} \label{eq-defS}
\end{align}

Each running instance (source), I is characterized by four attributes: CPU, Memory, Type and Cost, respectively denoted by:
\begin{align}
    I_i = \{P_i,R_i,T_i,C_i\}, \; \forall I \in S, \; i \in [1,2,...,M] \label{eq-defI}
\end{align}
 
 where M is the number of currently running instances. 
 
 $P_i$ is the average percentage CPU utilization plus two standard deviations (over a defined time period) received from CloudWatch, times the published Elastic Compute Units (ECUs) of instance $I_i$. 
 
 $R_i$ is the average percentage memory utilization plus two standard deviations (over a defined time period) received from CloudWatch, times the published memory capacity of instance $I_i$.
 
 Just like the set of running instances, each potential destination instance is denoted as: 
\begin{align}
 D = \{J_1,J_2,...,J_N\} \label{eq-defT}
\end{align}

where N is the number of publicly available EC2 instances in AWS and J is:
\begin{align}
    J_j = \{P'_j,R'_j,T'_j,C'_j\}, \; \forall J \in D, \; j \in [1,2,...,N]  \label{eq-defJ}
\end{align}

In equation (\ref{eq-defJ}), $P'_j$, $R'_j$ and $C'_j$ are the published CPU capacity (Elastic Compute Units), memory capacity and on-demand hourly cost of instance $J_j$.

The output of our model is a matrix that represents the recommended assignment for each of the running instances by EC2 instance type. This is denoted by: 
\begin{align}
 \text{Variable:} && X_{ij}, \; i = \{1,2,...,M\}, j = \{1,2,...,N\}
 \label{eq-Variable}
\end{align}

$X_{ij}$ is an M x N (source by destination) boolean matrix where a value of 1 means that the i-th running instance should be of type j while a value of zero means otherwise. This can be represented as \texttt{IF $X_{ij}$ == 1 THEN $T_i$ = $T'_j$}.
\begin{align}
 \text{Minimize:} && \sum_{i=1}^{m}\sum_{j=1}^{n} C'_j X_{ij} \label{eq-Min}
\end{align}

Equation (\ref{eq-Min}) represents the objective function of the optimization. It denotes that we would like to minimize the cost associated with each instance allocation. 

\begin{align}
 \left.\begin{aligned}
    X_{ij} \cdot P_i \cdot \partial_i \le P'_j \\
    X_{ij} \cdot R_i \cdot \partial_i \le R'_j
       \end{aligned}
 \right\}
 \qquad \; \partial_i \ge 1 \label{eq-con1}
 \end{align}
 
The goal of our model is to recommend instance types that are able to serve the resource demands of our set of running instances. To accommodate the potential for fluctuations in the expected workload, our model takes into consideration a user provided utilization factor ($\partial_i$) for each running instance. A value of 1 for $\partial_i$ suggests that future workloads will be similar to current or past workloads. However, since we don't typically desire full resource utilization, $\partial_i$ will normally be set at a value greater than or equal to 1 (see equation \ref{eq-con1}). On the other end, the larger the value of $\partial_i$, the higher the likelihood of inefficiency in resource assignment.  

\begin{align}
 \sum_{j=1}^{n} X_{ij} = 1, \forall i \in S \label{eq-con2}
\end{align}

The constraint represented in equation (\ref{eq-con2}), limits the recommendation of our model such that each running instance can only be assigned to one instance type. 
 
\subsection{AMPL Model}
We make use of AMPL \cite{ampl} to implement the algebraic model described in the previous section. AMPL provides us with a modeling language to describe, gather and manipulate our data; describe our variables, objectives and constraints; launch the optimization solver and analyze and output our results.

\begingroup
\fontsize{8pt}{10pt}\selectfont
\begin{verbatim}
1   set SERV;   #Servers
2   set INST;   #Instances
\end{verbatim} 
\endgroup

Lines \texttt{(1)} and \texttt{(2)} describe the set of currently running servers \texttt{SERV} and the set of instance types \texttt{INST} they are to be assigned to. 

\begingroup
\fontsize{8pt}{10pt}\selectfont
\begin{verbatim}
3   param cpu_s {INST} >= 0;   
4   param mem_s {INST} >= 0;   
5   param cpu_d {SERV} >= 0;   
6   param mem_d {SERV} >= 0;   
\end{verbatim} 
\endgroup

Lines \texttt{(3-6)} describe the resource supply (\texttt{cpu\_s, mem\_s}) and resource demand (\texttt{cpu\_d, mem\_d}) on the source and target respectively.

\begingroup
\fontsize{8pt}{10pt}\selectfont
\begin{verbatim}
7  param cost {SERV,INST} >= 0;
8  var Trans {SERV,INST} >= 0 , integer ;    
\end{verbatim} 
\endgroup
Lines \texttt{(7)} and \texttt{(8)} are M x N matrices representing the parameter for on-demand hourly cost of reach running instance by instance type and the boolean variable representing the assignment of each running instance by instance type respectively. Line \texttt{(8)} is the AMPL representation of the variable described by equation (\ref{eq-Variable}) in the algebraic formulation section above.

\begingroup
\fontsize{8pt}{10pt}\selectfont
\begin{verbatim}
9   minimize Total_Cost:
10      sum {i in SERV, j in INST} 
11          cost[i,j] * Trans[i,j];
\end{verbatim} 
\endgroup
Lines \texttt{(9-11)} is our objective function as described in equation (\ref{eq-Min}).

\begingroup
\fontsize{8pt}{10pt}\selectfont
\begin{verbatim}
12  subject to CPU{i in SERV, j in INST}:
13      Trans[i,j] * cpu_d[i] * d[i] <= cpu_s[j];
14
15  subject to Memory{i in SERV, j in INST}:
16      Trans[i,j] * mem_d[i] * d[i] <= mem_s[j];
17
18  subject to Total{i in SERV}:
19      sum {j in INST} 
20          Trans[i,j] = 1;
\end{verbatim} 
\endgroup
Lines \texttt{(12-20)} represent the model constraints described in equations (\ref{eq-con1}) and (\ref{eq-con2}), which state that the resource supply must be greater than or equal to the resource demand and that each running instance can only be assigned to one instance type.

\begin{figure}[h!]
\centering
\begingroup
\fontsize{8pt}{10pt}\selectfont
\begin{verbatim}
1  reset;
2
3  option show_stats 1;
4  option presolve 0;
5  option solver cplex;
6  option cplex_options 'display=2 presolve=0';
7
8  model aws.mod;
9
10 set CASES = 1 .. 31;
11 for {j in CASES} 
12   {
13      reset data;
14      data ("aws-" & j & ".dat");
15      solve;
16
17      option omit_zero_rows 1;
18      option omit_zero_cols 1;
19      option display_1col 200;
20        	        
21      display Trans > ("aws-" & j & ".out");
22      display Total_Cost >> ("aws-" & j & ".out");
23   }
\end{verbatim} 
\endgroup
\caption{AMPL ``aws.run'' file for our optimization problem.}
\label{fig:ampl-run}
\end{figure}

To execute the AMPL program for our optimization problem, we make use of the ``aws.run'' file in figure \ref{fig:ampl-run}. Because our problem is focused on the boolean assignment of running instances to instance types, we used the the CPLEX solver (see line \texttt{(6)}), which is suited for linear and quadratic optimization in integer and continuous variables \cite{CPLEX}. In order to evaluate the impact of the user provided utilization factor ($\partial_i$) on our optimization results, we generated 31 different data files for utilization factors of 1.0 to 4.0 in 0.1 increments. Lines \texttt{(10)} to \texttt{(23)} of our run file direct AMPL to iterate through all 31 data files, solve the optimization with the model file specified in line \texttt{(8)} and output the results to 31 different output files.

\section{Results}
\subsection{Cost}
\begin{figure}[h!]
\centering
\scalebox{.23}{
\includegraphics{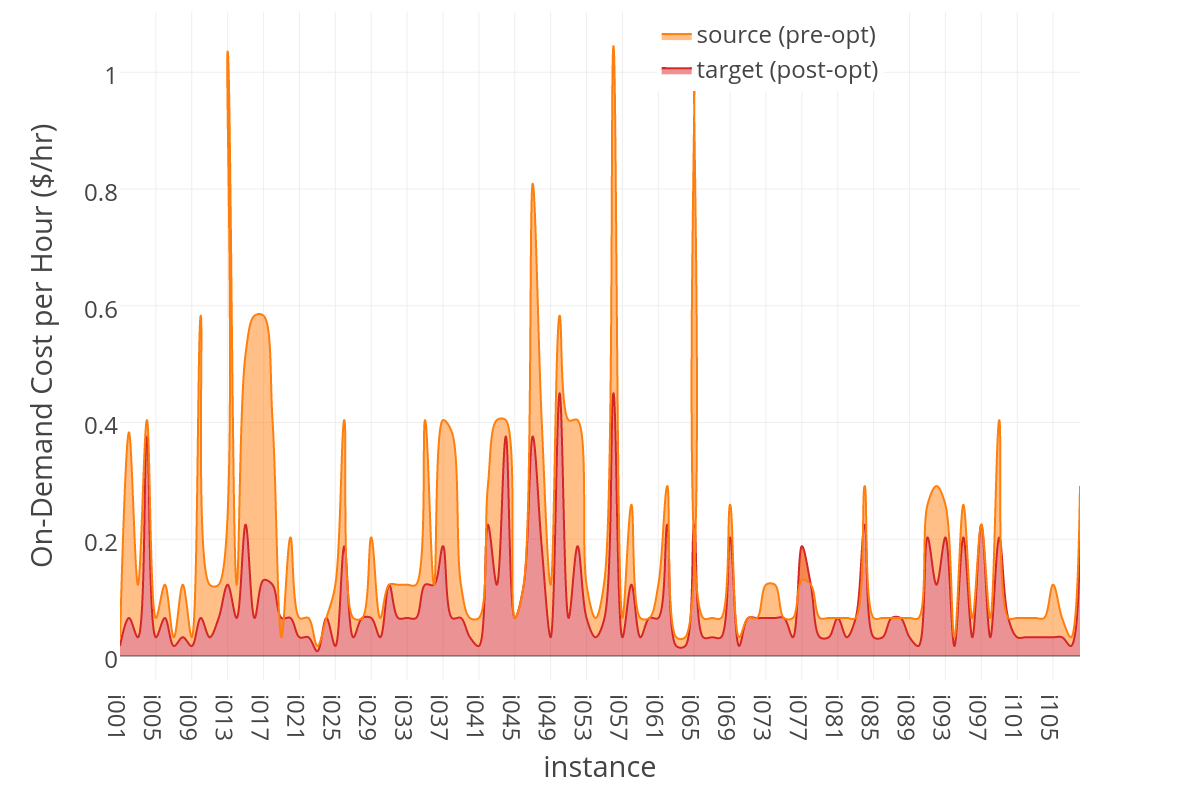}}
\caption{``Source (heuristic approach)'' versus ``Target (optimized approach)'' on-demand cost per hour (\$/hr) for each instance ($\partial_i = 1.5$ for all $I_i$).}
\label{fig:cost-source-target}
\end{figure}

With the utilization factor ($\partial_i$) for each running instance set at 1.5 (which implies that we expect our future workload demand to be 50\% more than our current workload demand), our results show an overall reduction in the per hour on-demand cost for each of the instances (see figure \ref{fig:cost-source-target}). The total hourly on-demand cost for all 108 instances which were considered in our study was \$21.09. This represents an annual cost of \$184,748 with the heuristic approach. At $\partial_i$ = 1.5, the projected total hourly on-demand cost for the target instances with the optimized approach is \$10.15. This represents a 52\% reduction in projected annual costs to \$88,914 when using the optimized approach. 

\vspace*{-2mm}

It is useful to note that not all instances experience cost reduction. While the overwhelming majority of the recommendations by our optimization algorithm is to size down in order to maintain the performance constraints on the model, our optimization algorithm does recommend that some of the instances be up-sized or that a different instance family be used. We will discuss the recommendations for instance family change in more detail later in the paper. 
\begin{figure}[h!]
\centering
\scalebox{.23}{
\includegraphics{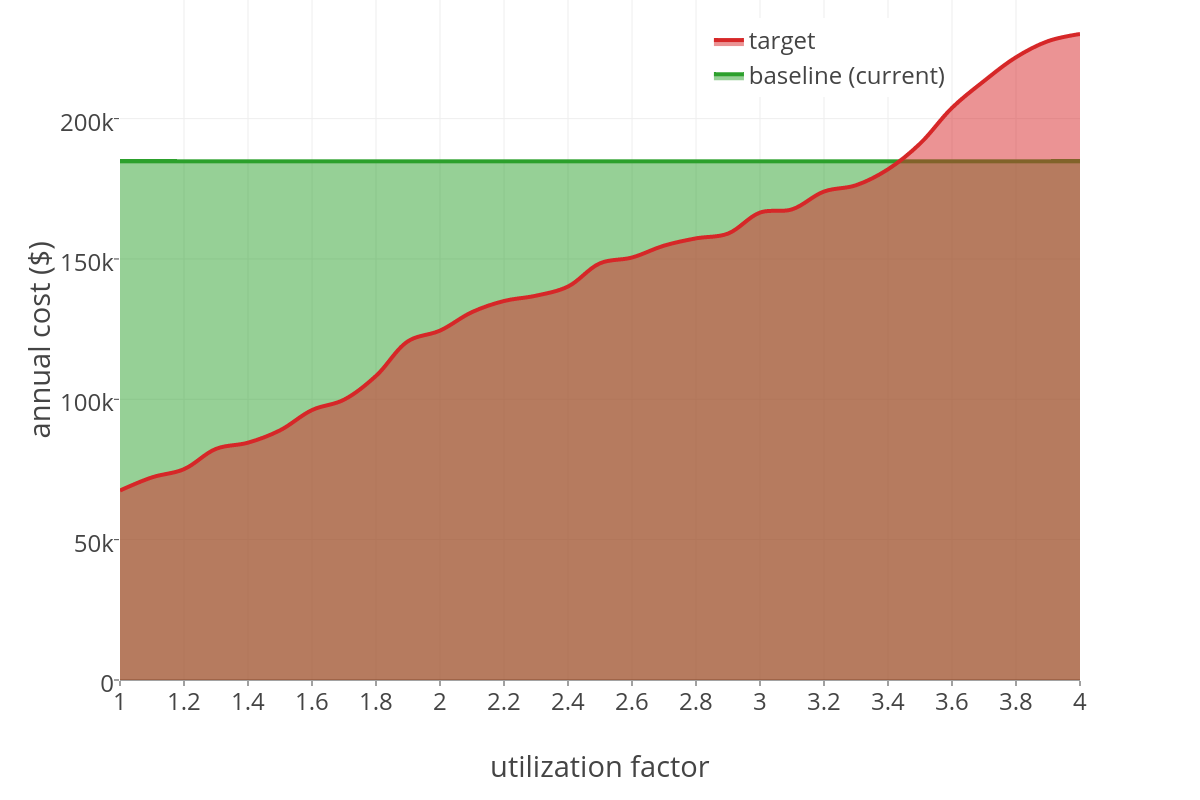}}
\caption{Baseline (current) annual cost versus target annual cost (\$) for $\partial_i = \{1,1.1,1.2,1.3,...,4\}, \; i \in [1,2,...,31]$.}
\label{fig:annual-cost-savings}
\end{figure}

In figure \ref{fig:cost-source-target}, we show the cost differences for each instance, pre- and post- optimization when $\partial_i$ is set to 1.5. Our results show a correlation between the value of $\partial_i$ and the hourly on-demand cost of each target instance. As we can see from figure \ref{fig:annual-cost-savings}, the projected annual cost for all 108 instances in our study increases as $\partial_i$ increases. This increase represents a corresponding reduction in the target savings (green zone) until the point when the projected annual costs exceed  (red zone) the current (baseline) annual cost of \$184,783. This break even point occurs between a $\partial_i$ of 3.4 and 3.5. Therefore, for the instances in our study, our model is best suited for optimizations when we anticipate future workload demand to be less than or equal to 3.4 times our current workload demand.
\begin{figure}[h!]
\centering
\scalebox{.23}{
\includegraphics{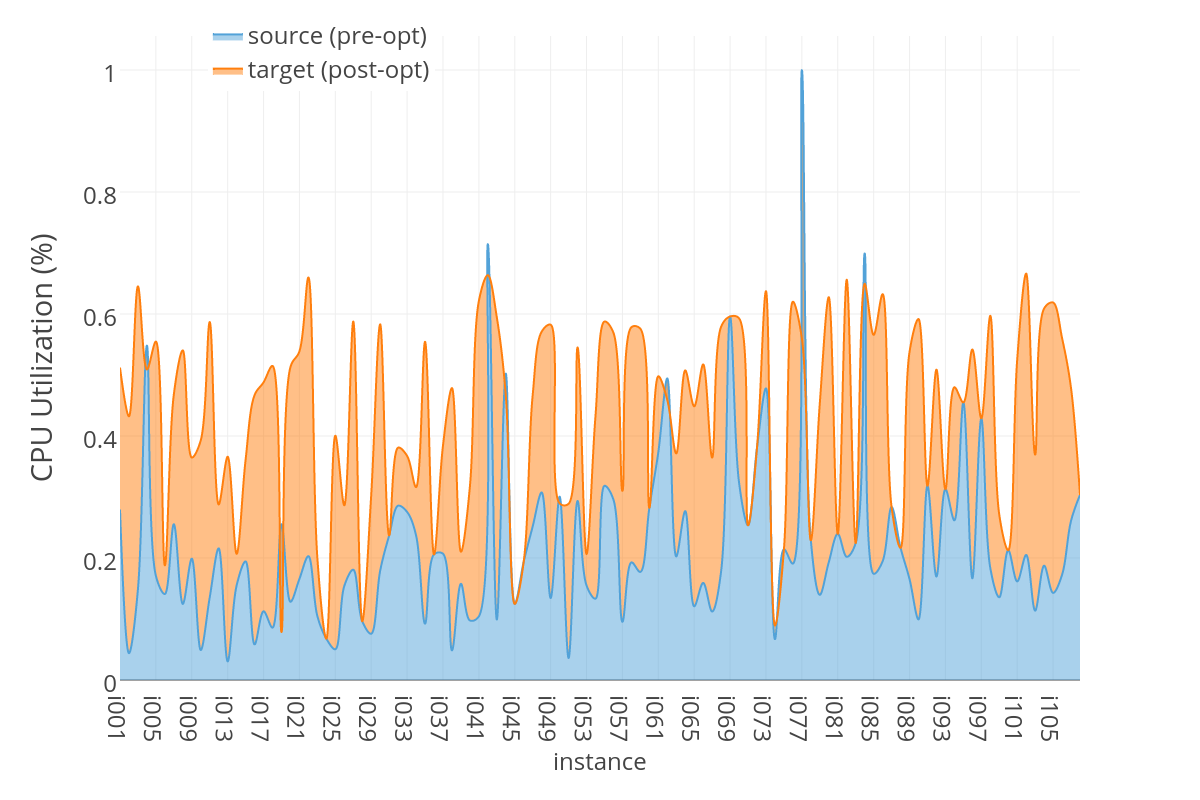}}
\caption{``Source (heuristic approach)'' versus ``Target (optimized approach)'' CPU utilization (\%) per instance ($\partial_i = 1.5$ for all $I_i$; source\_average $= 0.2236$; target\_average $= 0.4302$).}
\label{fig:cpu-util-source-target}
\end{figure}
\vspace*{-2mm}

\subsection{Resource Utilization}
\begin{figure}[h!]
\centering
\scalebox{.23}{
\includegraphics{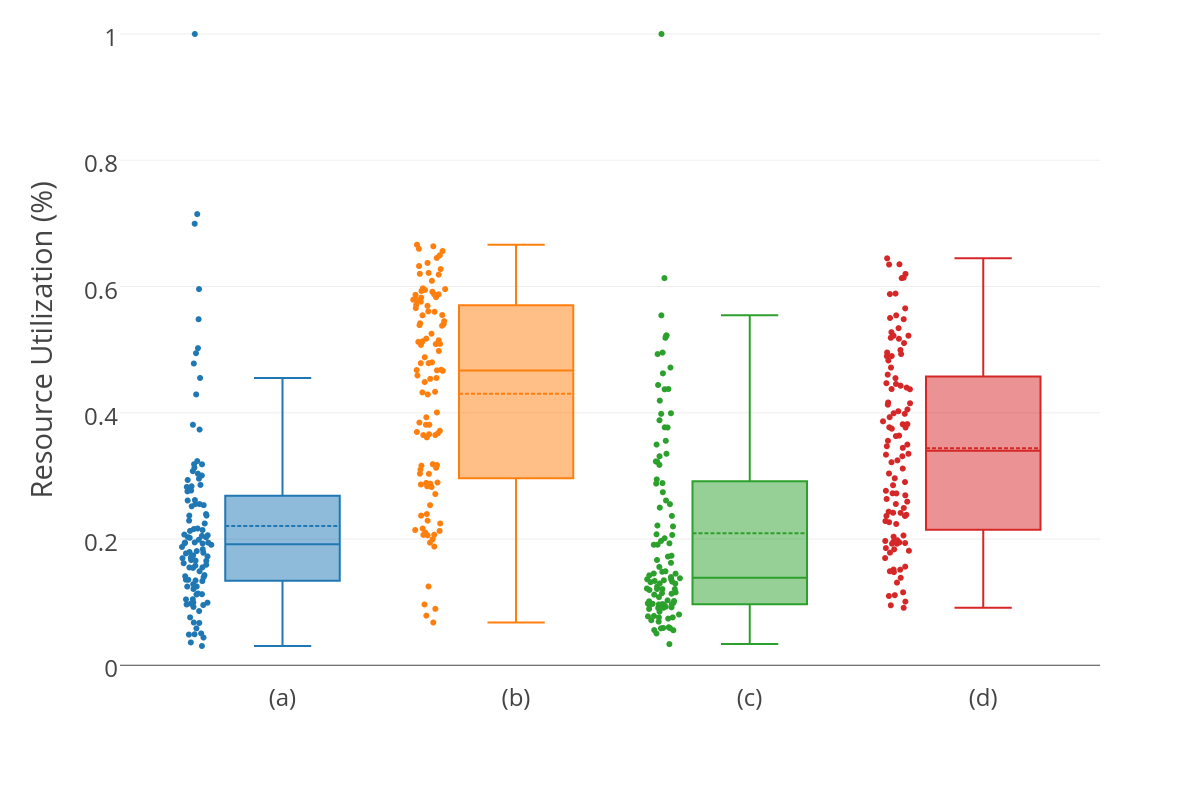}}
\caption{Distribution of resource utilization between the source instances and the target instances. (a) Source CPU utilization; (b) Target CPU utilization; (c) Source memory utilization; (d) Target memory utilization ($\partial_i = 1.5$ for all $I_i$).}
\label{fig:cpu-mem-util-box}
\end{figure}

While the stated and explicit objective of our model is to minimize cost, our model's implicit objective is to maximize resource utilization within the bounds of the defined constraints. Our results show an overall improvement in CPU utilization between the original running instances and the new instances as recommended by our model (see figure \ref{fig:cpu-util-source-target}). As figure \ref{fig:cpu-mem-util-box} (a, b) shows, average CPU utilization improves from 22.36\% to 43.02\% between the source and target instances. This mean increase is statistically significant at p $<$ .05,  with a t-value of -9.31628 and a p-value $<$ .00001 (two-tailed t-test).

\begin{figure}[h!]
\centering
\scalebox{.24}{
\includegraphics{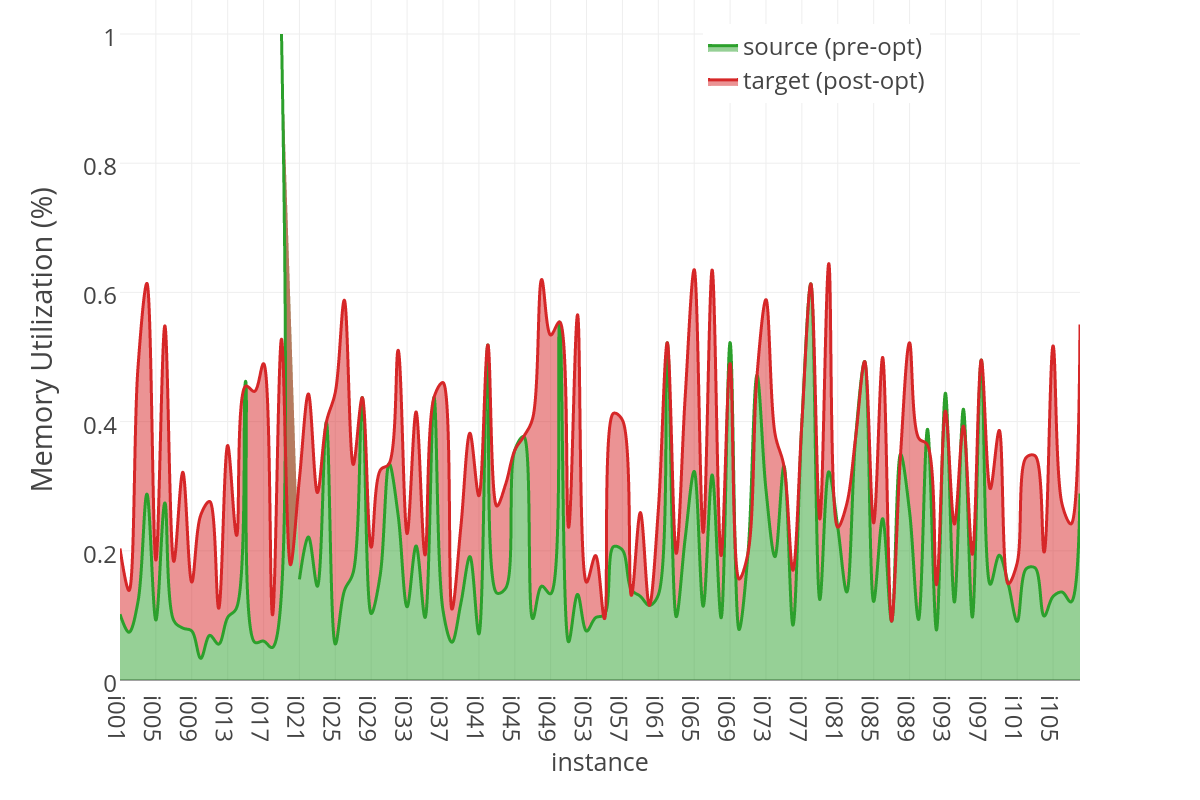}}
\caption{``Source (heuristic approach)'' versus ``Target (optimized approach)'' memory utilization (\%) per instance ($\partial_i = 1.5$ for all $I_i$for all $I_i$; source\_average $= 0.2096$; target\_average $= 0.3438$).}
\label{fig:mem-util-source-target}
\end{figure}

We also see an improvement in memory utilization between the original running instances and the new instances (see figure \ref{fig:mem-util-source-target}). The memory utilization mean improves from 20.96\% to 34.38\% as figure \ref{fig:cpu-mem-util-box} (c, d) shows. The difference in means is statistically significant at p $<$ .01,  with a t-value of -6.30688 and a p-value $<$ .00001 (two-tailed t-test).

\subsection{Instance Type Consolidation}
As mentioned in the previous section, the stated objective for our model is cost minimization within the bounds of the specified constraints. However, our results show that instance type consolidation between the pool of source instances and the pool of target instances is also a collateral benefit of our optimization model. 

\begin{figure}[h!]
\centering
\scalebox{.35}{
\includegraphics{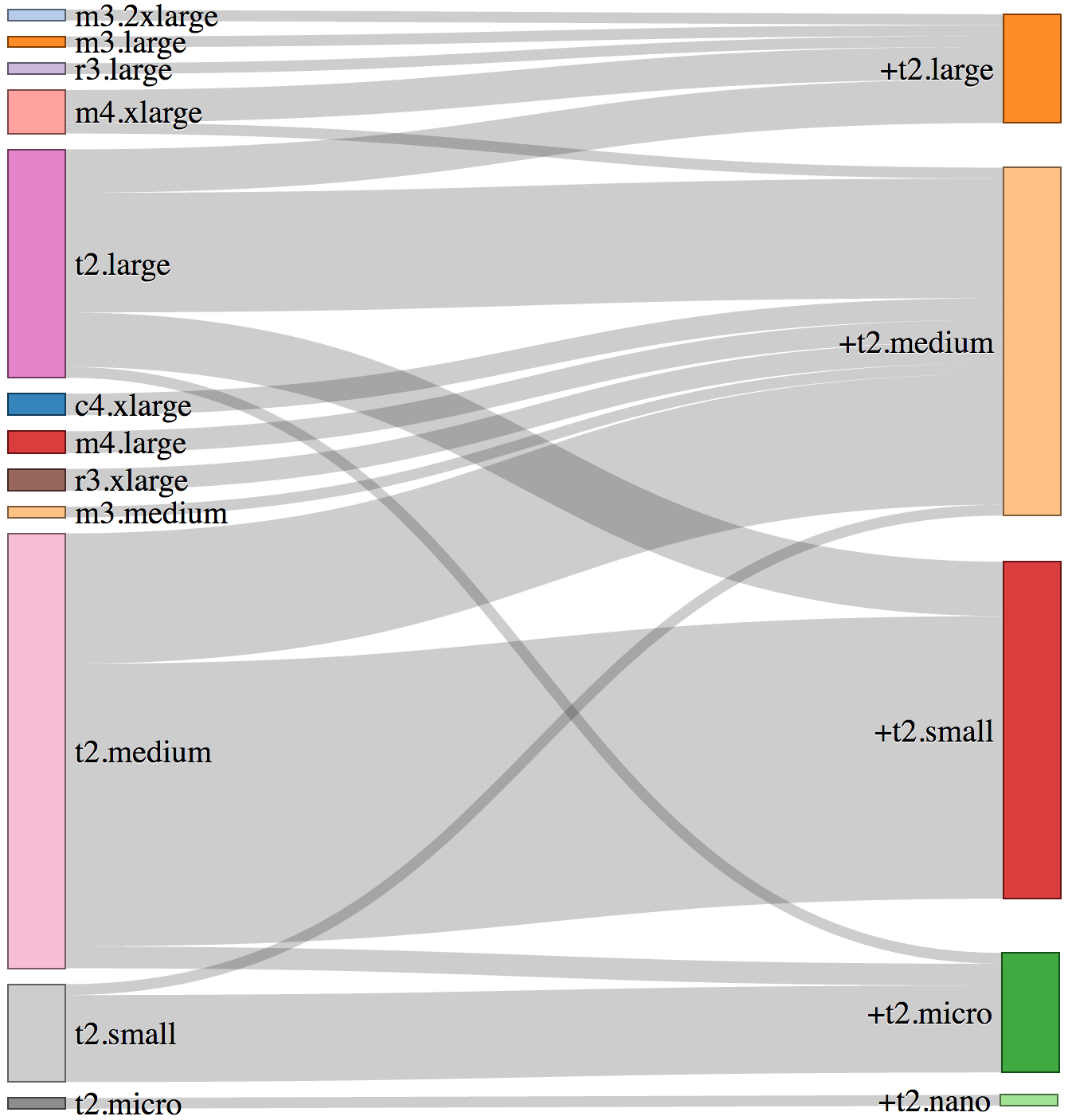}}
\caption{Instance type consolidation for t2-family of instances. The left column represents pre-optimization (source) instance types and the right column represents post-optimization (target) instance types ($\partial_i = 1.5$ for all $I_i$).}
\label{fig:t2-instance-flow}
\end{figure}

With $\partial_i$ set at 1.5, we see a reduction in the number of instance types between the source and target sets. The consolidation is more pronounced when considering the flow for the t2-family of instances. As we can see in figure \ref{fig:t2-instance-flow}, the source set of instances is made up of 12 distinct instance types, while the target set of instances is made up of just 5.

Instance type consolidation presents an opportunity for additional cost optimization when reserved instances are factored in. Amazon EC2 reserved instances provide a significant discount (up to 75\%) compared to on-demand instances. Reserved instance discounts are automatically applied to EC2 running instances when the attributes of the EC2 instance match the attributes of an active Reserved Instance. The attributes of a Reserved Instance are: instance type, platform, tenancy and availability zone \cite{aws-ec2-purchasing}. Therefore, the more consolidated instance types are, the more we can maximize the cost benefits of Reserved Instances.

\section{Conclusion}
Our work is based on metrics gathered from actual production workloads running in an Amazon Virtual Private Cloud. It highlights the benefits of a deductive approach over heuristics when allocating workloads on cloud infrastructure. Using an Integer Linear Programming (ILP) model for our optimization, we showed that for the given pool of workloads and Amazon EC2 instances analyzed, we can maximize resource utilization and minimize cost for projected future workloads up to 3.4 times the current demand levels. Our results also showed that we can optimize on instance types by consolidating towards the t2 family of instances.

\section{Future Work}
For future  work, we intend to take a look at the use of computer-based ranging \cite{chinneck-2000} for sensitivity analysis against the optimal solutions. By looking at how the solutions change when the right-hand side vectors and objective vectors are varied, we can better assess how much tolerance the fundamental solution has to changes to certain constraints of the linear programming model. 

In order to simplify our model, we only took into account CPU and memory metrics for our instances. A more comprehensive model will need to factor in the impact of I/O in terms of the EC2 instance (network) as well as the attached block storage volumes. To accommodate this, our model for $I_i$ and $J_j$ will need to be expanded to include the tuples $\{E_i,E'_j\}$ and $\{K_i,K'_j\}$ which represent the attributes for EC2 instance I/O and EBS disk I/O respectively. We would also need to make use of machine learning techniques for historical signal pattern matching and prediction in order to augment the value of the user provided utilization factor ($\partial_i$). 

\section*{Acknowledgment}
The authors would like to thank Brian Perri and Milind Saraph from the Office of Information Technologies at the University of Notre Dame for their assistance in extracting and curating the AWS CloudWatch metrics that were used in this research. 

\ifCLASSOPTIONcaptionsoff
  \newpage
\fi

\bibliographystyle{IEEEtran}
\bibliography{bibtex/biblio.bib}{}






\end{document}